\documentstyle[prl,aps, 12pt]{revtex}

\title{Directed polymers in a random medium: a variational approach}
\author{T. Garel ~and ~H. Orland \\
Service de Physique Th\'eorique\\
C.E.A.-Saclay, 91191 Gif-sur-Yvette Cedex\\
France}
\date{\today}

\def\be{\begin{equation}}
\def\th0{\theta_0}

\def\bea{\begin{array}}
\def\ee{\end{equation}} 
\def\eea{\end{array}}

\newcommand{\ddk}{\frac{d^d k}{(2 \pi)^d}}

\begin{document}
\maketitle
\vskip 1cm
\begin{abstract}
A disorder-dependent Gaussian variational approach is
applied to the problem of a $d$ dimensional polymer chain in a random
medium (or potential). Two classes of variational solutions are
obtained. For $d<2$, these two classes may be interpreted as
domain and domain wall. The critical exponent $\nu$
describing the polymer width is $\nu={1\over (4-d)}$
(domain solution) or $\nu={3\over (d+4)}$ (domain wall solution). The
domain wall solution is equivalent to the (full) replica symmetry
breaking variational result. For $d>2$, we find $\nu={1\over 2}$. No evidence
of a phase transition is found for $2< d< 4$: one of the variational solutions suggests that
the polymer chain breaks into Imry-Ma segments, whose probability distribution is calculated.
For $d>4$, the other variational solution undergoes a phase
transition, which has some similarity with B. Derrida's random energy models.
\end{abstract}
\vskip 3cm\noindent\mbox{Submitted for publication to:} \hfill 
\mbox{Saclay, SPhT/96-022}\\ \noindent \mbox{``Physical Review B'', Rapid Comm.}\\ 
\vskip 1cm\noindent \mbox{PACS: 75.10N, 64.70P, 71.55J} \newpage

\newpage

\section{Introduction}
\label{sec: intro}
Usually, in quenched disordered systems, extensive thermodynamical quantities,
such as the free energy, are identified to their average over 
the disorder\cite{Brout}. This can be understood in the
following way: one divides the macroscopic system into mesoscopic
subsystems, each subsystem corresponding to a particular
disorder configuration. For short-range forces, the free energy is
additive, and thus the total free energy is the sum of the free
energies of the subsystems. This line of reasoning is then implemented by the
use of the replica technique, in an exact way for mean field models 
\cite{Me_Pa_Vi} or
in a variational way for more realistic systems
\cite{Ed_Mu,Shak_Gu,Me_Pa}. 
The replica technique has been mostly tested 
for mean field models \cite{De_Ga1,Co_De} and references therein.
 
Recently, a disorder dependent variational approach has been
proposed for a large class of quenched disordered systems
\cite{Orland_Shapir}. Roughly speaking, this method does the converse
of replica variational methods: one looks for the best
translationnally invariant Hamiltonian, for a fixed disorder configuration.
One major advantage of this 
approach is that it is genuinely variational, thus
providing a true upper bound to the
free energy of the system, unlike replica based methods, which are
``plagued by the $n=0$ limit''. In fact, this method is closely
related to the variational replica method, as shown for the random
sine-Gordon \cite{Orland_Shapir} or the random-field XY model\cite{Ga_Io_Or}.
In this paper, we apply this method to the problem of a $d$
dimensional directed polymer chain in a random medium (or potential).
This problem is related to many non-linear and/or disordered systems,
such as random manifolds \cite{Bale_Bou_Me}, kinetic growth
\cite{Hal_Zha}, turbulence \cite{Bou_Me_Pa}, and we hope that the
present approach can shed some light on these problems.
The Hamiltonian of the chain reads (in the following $\beta={1\over
T}$, where $T$ is the temperature): 
\begin{equation}
\label{ham}
\beta{\cal H} = {1 \over {2a^2}} {\int_0^L ds} \ {\dot {\vec r}(s)}^2 + \beta \int_0^L
ds \ {V(\vec r(s),s)}
\end{equation}
where $a$ is some microscopic length, $\vec r(s)$ denotes the $d$ dimensional position of link $s$ of
the chain ($s \in [0,L]$), and $V(\vec r(s),s)$ denotes the random
potential acting on this link. In this paper, we will restrict ourselves to
a (site uncorrelated) Gaussian potential, the correlation function of
which reads:
\begin{equation}
\label{proba}
\overline {V(\vec r,s) V(\vec r',s')}= V^{2} \ \delta(\vec r-\vec r')\delta(s-s')
\end{equation}
where $\overline A$ denotes the average of a quantity $A$ over the disorder. 

The layout of this paper is the following. In section II, we define the 
variational Hamiltonian, and calculate the corresponding variational
free energy. Two different classes of variational solutions are
exhibited. In section III, 
we discuss in detail the case of space dimension $d<2$, where these
two classes yield two different polymer width exponent $\nu$.
In section IV, we study the case $d>2$, , where we get an exponent
$\nu= {1 \over 2}$ for all solutions. We argue that for $d>4$, one of the
variational solutions points undergoes a phase transition which is not unlike
the (simple or generalized) random energy model. No evidence for a
phase transition is found for $2 < d <4$ : one of the variational
solutions can be interpreted in terms of Imry-Ma segments. In the
conclusion, we discuss the implications of these results, and suggest that a
renormalization procedure should be performed together with this
variational method.

\section{The variational free energy }
\label{sec: free}
We consider Hamiltonian (\ref{ham}) and its associated variational Gaussian companion:
\be
\label{h0}
\beta {\cal H}_0 = {1\over 2}{\int_0^L ds}{\int_0^L ds'}  \ (\vec r(s)-\vec
R) \ g^{-1}(s-s') \ (\vec r(s')-\vec R)
\ee
where we have restricted the variational kernel $g$ to be
translationally invariant. Note that contrary to the variational
replica method where the disorder is a priori averaged, one has to
consider here the possibility of a variational shift
$\vec R$ for each link. (It would be even better to consider
a variational link-dependent parameter $\vec R(s)$, but the
calculation are too involved).
The true free energy
$F$ satisfies the bound:
\begin{equation}
\label{bound}
\beta F \le \beta \Phi (V) = \beta F_0 + \beta <{\cal H}-{\cal H}_0>_0
\end{equation}
where $<...>_0$ stands for the thermal average with Hamiltonian (\ref{h0}).
Using equations (\ref{ham}),(\ref{h0}), and (\ref{bound}), together with periodic boundary
conditions ${\vec r}(0) ={\vec r}(L)$, we obtain the disorder
dependent variational free energy as:
\be
\label{Phi}
\beta \Phi({V}) = - d \sum_{n=1}^{\infty} ~ \log {\tilde g_{n}\over a^{2}} +
d \sum_{n=1}^{\infty} \ L {\omega_{n}}^2 \ {\tilde g_{n}\over a^{2}} + \beta {\cal W}(\vec
R ,G)
\ee
with $\omega_{n}={2n\pi \over L}$ and
\be
\label{Delta}
{\cal W}(\vec R,G)=\int {\ddk} \int_0^L ds \ V(\vec k,s) \ e^{i{\vec
k}\cdot{\vec R}} \ e^ {-{G \over 2} \ {\vec k}^2}
\ee
and
\be
\label{Ge}
G= 2 \sum_{n=1}^{\infty}\tilde g_{n}=2 \sum_{n=1}^{\infty} \left(\int_0^L ds \ g(s) \ e^{i\omega_{n} s}\right)
\ee
Since $\vec R$ is independent of $s$, and using equation (\ref{proba}), we get
\be
\label{def1}
{\cal W}(\vec R,G)=L ^{1\over 2} \int {\ddk} \ U(\vec k) \ e^{i{\vec
k}\cdot{\vec R}} \ e^ {-{G \over 2} \ {\vec k}^2}
\ee
with $\overline {U(\vec k)U(\vec k')}=V^{2} \delta (\vec k+\vec k')$.

\subsection{The variational solutions}
\label{solutions}
The minimization equations with respect to $\tilde g_{n}$ and $\vec R$ read:
\be
\label{gn}
\tilde g_{n}=  a^{2} \ {1\over {L {\omega_{n}}^2 +{\beta a^{2}\over d} {\nabla}^2 _{\vec R}{\cal W}(\vec R,G)}}
\ee
and 
\be
\label{min}
\vec {\nabla}_{\vec R} {\cal W}(\vec R,G) =\vec {0}
\ee
Note that one may find several solutions to the variational equations.
In disordered systems, one does not expect, in general, that their
(variational) free energy differ by an extensive amount. Since
fluctuations around one solution, or instantons connecting different
solutions, may yield extensive contributions to the free energy, one
has a priori to keep all variational solutions, unless some can be shown
to be unstable with respect to such fluctuations \cite{Ga_Io_Or}. In
the present problem,  we first point out that there are two classes of
solutions as is clear from equation (\ref{gn}):

(i) the first, hereafter denoted by $(+)$ has  ${\nabla}^2 _{\vec
R}{\cal W}(\vec R_{+},G_{+}) \ > \ 0$. 

(ii) the other, denoted by $(-)$, has  ${\nabla}^2 _{\vec R}{\cal
W}(\vec R_{-},G_{-}) \ < \ 0$. 

Defining $\alpha_{+}={\beta L a^{2}\over {4\pi^2 d}}{\nabla}^2 _{\vec R}{\cal
W}(\vec R_{+},G_{+})$, equation (\ref{Ge}) can be rewritten
\be
\label{red}
G_{+}= {L a^{2}\over 2\pi^2} \sum_{n=1}^{\infty}{1\over {n^2 + {\alpha_{+}}}}
\ee
 for the ($+$) solution. In a similar way, we define
$\alpha_{-}={\beta L a^{2}\over {4\pi^2 d}}\vert{\nabla}^2 _{\vec R}{\cal
W}(\vec R_{-},G_{-})\vert$, and equation (\ref{Ge}) can be rewritten  
\be
\label{green}
G_{-}= {L a^{2}\over 2\pi^2} \sum_{n=1}^{\infty}{1\over {n^2 - {\alpha_{-}}}}
\ee
for the ($-$) solution. Note that we must have  $\alpha_{-} \le 1$ for 
stability reasons, since equation (\ref {h0}) requires that $\tilde g_n >
0, \ \forall n$. 
Using equations (1.421.3 and 1.421.4) of reference
\cite{Grad_Ry}, we may rewrite equations (\ref{red}) and (\ref{green}) as: 

\be
\label{grad1}
G_{+}={La^{2} \over {4\pi}} \ {1\over
{\sqrt{\alpha_{+}}}}(\coth({\pi\sqrt{\alpha_{+}}})-{{1\over
\pi\sqrt{\alpha_{+}}}})
\ee
and
\be
\label{grad2}
G_{-}=-{La^{2} \over {4\pi}} \ {1\over
{\sqrt{\alpha_{-}}}}(\cot({\pi\sqrt{\alpha_{-}}})-{{1\over \pi\sqrt{\alpha_{-}}}})
\ee

The variational free energies $\Phi_{\pm}(V)$ are easily obtained through
equation (\ref{Phi}). Denoting $\Psi_{\pm}(V)=\Phi_{\pm}(V)-\Phi(0)$,
we get
\be
\label{Psip}
\beta \Psi_{+}=\beta ({\cal W}(\vec R_{+},G_{+}) -{1\over 2} G_{+} {\nabla}^2 _{\vec R}{\cal
W}(\vec R_{+},G_{+}))+\log ({\sinh ({\pi\sqrt{\alpha_{+}}})\over
{\pi\sqrt{\alpha_{+}}}})
\ee
 for the $({+})$ solution and 
\be
\label{Psim}
\beta \Psi_{-}=\beta ({\cal W}(\vec R_{-},G_{-})+{1\over 2} G_{-} \vert{\nabla}^2 _{\vec R}{\cal
W}(\vec R_{-},G_{-})\vert)+\log ({\sin ({\pi\sqrt{\alpha_{-}}})\over
{\pi\sqrt{\alpha_{-}}}})
\ee
for the $({-})$ solution.

\subsection{On the number of solutions and typical values of random quantities}
\label{remark}
In principle, one should solve the minimization equations to find
the disorder dependent quantities of interest $G_{\pm}$ and $R_{\pm}$.
In practice, we will estimate typical orders of magnitude in the
following way. For instance equation (\ref{def1}) implies
\be
\overline {{\cal W}(\vec R_1,G){\cal W}(\vec R_2,G)}=L \overline{
\int {\frac{d^d k_1}{(2 \pi)^d}}\int {\frac{d^d k_2}{(2 \pi)^d}} \ U(\vec k_1) \ U(\vec k_2) \ e^{i{\vec 
k_1}\cdot{\vec R_1}+i{\vec 
k_2}\cdot{\vec R_2}} \ e^ {-{G \over 2} \ ({\vec k_1}^2+{\vec k_2}^2)}}
\ee
Clearly, we are only able to estimate this quantity if we temporarily
forget that $G$ and $\vec R$ depend themselves on $U(\vec k)$.
This decoupling procedure then yields
\be
\label{corr}
\overline {{\cal W}(\vec R_1,G){\cal W}(\vec R_2,G)} \simeq {LV^2 \over (2{\sqrt
\pi}G)^{d\over 2}}   \ e^{-{(\vec R_1-\vec R_2)^2\over 4G}}
\ee
so that a typical value of ${\cal W}(\vec R,G)$ reads
\be
\label{order}
\left({\cal W}(\vec R,G)\right)_{typ} \simeq \ V \ {L^{1\over 2}\over G^{d\over 4}}
\ee
up to a random (algebraic) constant. In the same conditions, we obtain
\be
\label{ordre}
\left(\alpha_{\pm}\right)_{typ}  \simeq \ V \ {L^{3\over 2}\over G_{\pm}^{1+{d\over 4}}}
\ee
up to a random (positive) constant. 

To get a feeling for its range of validity, one may calculate, within
this approximation, the averaged number of points $\vec R$ that
satisfies the minimization equation (\ref{min}), or equivalently the
averaged (or typical) distance between such two such points $\vec
R_{1}$ and $\vec R_2$. A straightforward calculation, based on equations 
 (\ref{min}), and (\ref{corr}) shows that $\left(\vert\vec R_1-\vec R_2\vert\right) \sim \sqrt{G}$. 

We thus expect that the disorder dependent variational method, and the
approximate estimation of typical random quantities are justified if
there are few solutions, that is if ${G_{\pm}}$ is large. 

\subsection{Stability of the solutions}
\label{stability}
We will also examine the stability of the variational solutions with
respect to the variational parameters $\tilde g_{n}$ and $\vec R$. As
explained in reference \cite{Ga_Io_Or} for the random field XY
model, we do not expect any instability with respect to the $\tilde
g_{n}$'s and consider only the stability of the solution $(G,\vec R)
$ with respect to small link-dependent shifts $\vec {\delta}(s)$. 
The associated free energy reads
\be
\label{fluct}
\Delta (\beta\Phi) = \int ds \left({1\over 2a^2}\dot {\vec
{\delta}}^{2}(s)-\beta\int {\ddk}  \ (\vec k\cdot\vec
{\delta})^{2} \ V(\vec k,s) \ e^{i{\vec
k}\cdot{\vec R}} \ e^ {-{G \over 2} \ {\vec k}^2}\right)
\ee 
The positivity of the $\Delta (\beta\Phi)$ is determined by the
spectrum of the kernel in (\ref{fluct}). This kernel is analogous to
that of a Schrodinger equation in a random potential of typical
strength $\left({V\over {G^{1+{d\over 4}}}}\right)$. For large values of $G$, the
potential vanishes, leading to marginal (zero energy) fluctuations.

\section{Results for $\protect{d<2}$}
For physical purposes, this means essentially $d=1$. In this case, one
has an exact solution \cite{Kard} with $\nu={2\over 3}$ (and corrections to the free energy of
order $L^{1\over 3}$), and a variational replica calculation
\cite{Me_Pa} with $\nu={3\over 5}$ (and
corrections to the free energy of order $L^{1\over 5}$). We now
consider the two classes of variational solutions, which (in $d=1$) can
be called potential minima or maxima. 
\par
\subsection{The $({+})$ solution} 
It is easily checked, using (\ref{grad1}), 
 that for $d=1$, the only self consistent solution
of equation (\ref{red}) is $\alpha_{+} \simeq L^{2\over 3}$ and 
\be
\label{plus}
G_{+} \simeq  \ L^{2\over 3}
\ee
that is an exponent $\nu={1\over 3}$, or more generally $\nu={1\over (4-d)}$
for $d<2$. As shown in section (\ref{stability}), this
solution is marginally stable. Moreover, its physical meaning can
be appreciated through a Flory argument; since the $({+})$ solution
corresponds to attractive (``collapsed'') regions, a balance between
the entropic  term and an Imry-Ma estimate \cite{Im_Ma} of the
potential term yields  
\be
\label{flory1}
{L\over G} \simeq \left({L\over G^{d\over 2}}\right)^{1\over 2}
\ee
which indeed yields $\nu={1\over (4-d)}$. For $d=1$, the disorder
dependent part of the variational
free energy $\Psi_{+}$ is of order $L^{1\over 3}$.
\subsection{The $({-})$ solution}
From equation (\ref{green}), one must have $\alpha_{-} \le 1$. This
implies for $d=1$ 
\be
\label{moins}
G_{-} \simeq L^{6\over 5}
\ee
yielding $\nu={3\over 5}$, or more generally $\nu={3\over (d+4)}$ for $d<2$.
This solution is also marginally stable. It
can also be obtained, in a Flory like manner, applied now to the
repulsive (or swollen) regions
\be
\label{flory2}
{G\over L} \simeq \left({L\over G^{d\over 2}}\right)^{1\over 2}
\ee
In this case, the disorder dependent part of the variational free energy (see equation (\ref{Psim}))
is of order $L^{1\over 5}$, in agreement with the Flory estimate (and the
variational replica result). 

These results strongly suggest that the $({-})$ solution is very similar
to the full symmetry breaking replica solution. The second length
scale which comes out of the $({+})$ (or domain) solution has not been
obtained by the other methods. Note that its free energy $\Psi_{+}$ is
of order $L^{1\over3}$. Since $G$ diverges with $L$ for both
solutions, we expect our variational approach to be meaningful: (i)
there are few such solutions (ii) these solutions are marginally stable.

\section{Results for $\protect{d>2}$}
The problems we face for $d>2$ are threefold:

(a) in our approximations, the disorder becomes almost irrelevant
for $L$ large and $G$ large. In marked contrast with the $d<2$ case,
equations (\ref{grad1},\ref{grad2}) give a single solution, $G \simeq L$,
together with $\alpha_{\pm} \simeq V L^{(2-d)\over 4} \simeq 0$. The
fact that the exponent $\nu$ sticks to its Brownian value above two
dimensions has been also obtained in the variational replica method 
\cite{Me_Pa}.

(b) the identity of solutions $({+})$ and $({-})$ does not survive if
we allow for  variational solutions where either $G$ or $L$, or
both, become finite. In this case, the
variational method we have used requires at least a new
interpretation, since it has many variational solutions (if $G$ is
finite), or considers only a finite portion of the chain
(if $L$ is finite), or both. It is also possible that such solutions
are unstable (see section (\ref{stability})).

(c) to make matters worse, most of the high dimension models 
deal with directed polymers on a lattice. Most prominent among these
lattice models are the three-dimensional and infinite-dimensional
(tree) models (\cite{Co_De} and references therein). It is clear that
the comparison of the continuous model described by equation
(\ref{ham}) with these discrete models is not obvious, notwithstanding
the very existence of the continuum limit \cite{Hal_Zha}. 

With all these caveats in mind, we will now discuss two particular solutions
of the variational equations, which may bridge the gap between
discrete and continuum models.

\subsection{The large $L$, small $G$, $({+})$ solution for $d>4$}

It is possible to find a solution $G_{+} \simeq 0$ for large $L$, as
seen from equation (\ref{grad1}) in the limit of large $\alpha_{+}$.
We get ${G_{+}\over a^2} \simeq L ^{2\over (4-d)}$ which indeed vanishes
for $d>4$. (This solution corresponds, for $d<2$, to the domain solution). 
Even though we do not wish to discuss in detail how the
limits $L$ large and $a$ small are to be taken, we will see below that
indeed $d=4$ is a borderline dimension.

The disorder dependent part of the variational free energy
(\ref{Psip}) reads for large $\alpha_{+}$ 
\be
\label{Psip1}
\Psi_{+} \simeq {\cal W}(\vec R_{+},G_{+}) -{1\over 2} G_{+} {\nabla}^2 _{\vec R}{\cal
W}(\vec R_{+},G_{+})+{1\over \beta}({\pi\sqrt{\alpha_{+}}})
\ee

The first term on the r.h.s. of equation (\ref{Psip1}) is to be evaluated with
the constraint that  ${\nabla}^2 _{\vec R}{\cal W}(\vec R_{+},G_{+})> 0$. 
It is then easily shown that the algebraic constant omitted in equation
 (\ref{order}) is negative. The second term on the r.h.s. of (\ref{Psip1}) is 
also negative. Finally, the third term is positive: since we have
$\Psi_{+}=\Phi_{+}(V)-\Phi(0)$, this implies the 
existence of a phase transition between a high temperature Brownian
phase ${G_{+}\over a^2}\simeq L$ and a low temperature ``frozen phase''
${G_{+}\over a^2}\simeq L^{2\over 4-d}$. Note that the free energy
of this frozen phase scale like $\sqrt{\alpha_{+}}$, i.e. like
$L^{(d-2)\over (d-4)}$, which explicitly shows the problems of the
continuum version of the model as compared to its lattice
counterpart.

As stressed above, this phase transition pertains to a single $(+)$
solution. The typical distance between two ``frozen'' solutions being
of the order of $a$, we are thus faced with an exponential number (in
$L$) of such solutions. To get a flavour of the nature of the phase
transition, we calculate a typical correlation between the free
energies of two such solutions, and obtain, within the decoupling
scheme of section (\ref{remark}): 
\be
\label{rem}
\overline {{\cal W}(\vec R_{1}, G_{+}){\cal W}(\vec R_{2}, G_{+})}
\simeq {LV^2\over ({2\sqrt\pi G_{+}})^{d\over 2}} \ e^ {-\left( {(\vec
R_{1}-\vec R_{2})^{2}\over 4{G_{+}}}\right)} \ \simeq
\pi^{d\over 2}LV^2
\ \delta(\vec R_{1}-\vec R_{2})
\ee
These correlations are indeed reminiscent of the (simple or
generalized) random energy model. We cannot evaluate 
the exponent $\nu$ in the low temperature phase, but we think it is
also $\nu={1\over 2}$, since the polymer undergoes a random walk
between variational frozen solutions, that is a random walk on the
different $(\vec R_{i})$ points.
We tentatively conclude that dimension $d=4$ may well be some kind of
lower critical dimension for a (simple or generalized) random energy
model type of phase transition. The role of
dimension $d=4$ in this context has been recently discussed in \cite{Bun_Las}.

\subsection{The finite $L$, finite $G$, $({-})$ solution for $d>2$}
Another intriguing result of the variational equations concerns the
$({-})$ solution, since it is restricted by the condition $\alpha_{-}
\le 1$. This condition does not play any role in the (large $L$, large
$G_{-}$) solution, for $d>2$. However, if one considers, a finite $G_{-}$
solution, it can only exist up to a number of links $L \le L_0$,
(see equation (\ref{green}) and the definition of $\alpha_{-}$) with
\be 
\label{IM}
{L_0}^{{3\over 2}}={4\pi^{2} d\over \beta}  \ \left( \int {\ddk} \  {\vec k}^{2} U(\vec k) \ e^{i{\vec
k}\cdot{\vec R_{-}}} \ e^ {-{G_{-}\over 2} \ {\vec k}^2} \right)^{-1}
\ee
Knowing the probability distribution of $U(\vec k)$, and using the
decoupling approximation of section (\ref{remark}), one may evaluate
the probability distribution of $L_0$ as
\be
P(L_0)=A(d,\beta V) \ \left({G_{-}^{1+{d\over 4}}\over L_0^{5\over 2}}\right) \
\exp-\left(C(d,\beta V){G_{-}^{2+{d\over 2}}\over L_0^{3}}\right)
\ee
where $A(d,\beta V)$ and $C(d,\beta V)$ are regular functions of the
dimension and of the temperature. Note that the second (and higher)
moment of this distribution is divergent. This result strongly
suggests that the chain breaks into Imry-Ma domains of (distributed)
size $L_0$. A more detailed description (such as the role of the scale $L_0$
in the overlap between different variational solutions) requires a better
understanding of the spatial succession of (${+}$) and (${-}$)
finite $L$ solutions along the chain.

\section{Conclusion}
In this paper, we have presented a disorder dependent variational
method for the problem of a $d$ dimensional directed polymer in a random
potential. This method seems reliable for $d<2$ , where there are few
variational solutions, and agrees when they overlap, with the (full)
replica symmetry breaking variational method. We have also found a
new length scale (the domain solution), which is apparently missed by
other approaches. For $d>2$, the variational solutions may be very dense
and our variational procedure should be viewed as a first
step towards a variational renormalization group: the
free energy $\Psi(\vec R, G)$ has indeed (see equations (\ref{Psip}),
(\ref{Psim})) the form of a new random potential, so one may think of
iterating the process \cite{Ga_Or}.
In this approach, we have presented some peculiar solutions
which may have some relevance, either to the puzzle of critical
dimensions for this problem, or to the physical description of the
chain. We have explicitly shown that $d=4$ plays a special role for
the $(+)$ solutions, and that an Imry-Ma length $L_0$ is, for $d>2$,
the natural scale for the correlations between different finite
$G_{-}$ solutions. A more ambitious goal would be to study the spatial
interplay of the $({+})$ and (${-})$ solutions in the variational 
renormalisation procedure to see if chaotic behaviour may arise \cite{Berk}.  

We thank J-P. Bouchaud and D. Thirumalai for useful comments and discussions. 

\newpage
{\centerline{\bf REFERENCES}}

\begin{references}
\bibitem{Brout}
R. Brout, Phys. Rev. {\bf 115}, 824 (1959).
\bibitem{Me_Pa_Vi}
M. M\'ezard, G. Parisi and M.A. Virasoro, \lq\lq Spin glass theory 
and beyond \rq\rq, World Scientific, Singapore, 1987.
\bibitem{Ed_Mu}
S.F. Edwards and M. Muthukumar,  J. Chem. Phys. {\bf 89}, 2435 (1988).
\bibitem{Shak_Gu}
E.I. Shakhnovich and A.M. Gutin,  J.Phys. {\bf A22}, 1647 (1989).
\bibitem{Me_Pa}
M. M\'ezard and G. Parisi, J. Phys. {\bf I}, 1, 809 (1991)
\bibitem{De_Ga1}
B. Derrida and E. Gardner, J.Phys. {\bf C19}, 2253, 5783 (1986).
\bibitem{Co_De}
J. Cook and B. Derrida, J. Stat. Phys. {\bf 63}, 505 (1991)
\bibitem{Orland_Shapir}
H. Orland and Y. Shapir, Europhys. Lett. {\bf 30}, 203 (1995).
\bibitem{Ga_Io_Or}
T. Garel, G. Iori and H. Orland, Phys. Rev. {\bf B 53}, R2941 (1996).
\bibitem{Bale_Bou_Me}
L. Balents, J-P. Bouchaud and M. M\'ezard, \lq\lq The large scale energy
landscape of randomly pinned objects \rq\rq, preprint cond-mat {\bf 9601137}.
\bibitem{Hal_Zha}
T. Halpin-Healy and Y.C. Zhang, Phys. Rep. {\bf 254}, 217 (1995)
\bibitem{Bou_Me_Pa}
J-P. Bouchaud, M. M\'ezard and G. Parisi, Phys. Rev {\bf E 52}, 3656 (1995).
\bibitem{Grad_Ry}
I.S. Gradshteyn and I.M. Ryzhik, \lq\lq Table of Integrals, Series,
and Products \rq\rq, Academic Press, New York, 1980.
\bibitem{Kard}
M. Kardar, Nucl. Phys., {\bf 290}, 582 (1987)
\bibitem{Im_Ma}
Y. Imry and S.-k. Ma, Phys. Rev. Lett. {\bf 35}, 1399 (1975).
\bibitem{Bun_Las}
R. Bundschuh and M. Lassig, \lq\lq Directed polymers in high 
dimensions \rq\rq, preprint cond-mat {\bf 9602045}.
\bibitem{Ga_Or}
T. Garel, H. Orland, in preparation.
\bibitem{Berk}
S.R. McKay, A.N. Berker and S. Kirkpatrick, Phys. Rev. Lett. {\bf 48},
767 (1982).







\end{references}


\end{document}